\documentclass{article}

\usepackage[numbers]{natbib}
\usepackage[utf8]{inputenc} 
\usepackage[T1]{fontenc}    
\usepackage{url}            
\usepackage{booktabs}       
\usepackage{amsfonts}       
\usepackage{nicefrac}       
\usepackage{microtype}      
\usepackage{xcolor}     
\usepackage{times}
\usepackage{graphicx}

\title{ML framework for global river flood predictions based on the Caravan dataset}

\author{
Ioanna Bouri\footnote{University of Helsinki, Finland.},
Manu Lahariya\footnote{University of Ghent, Belgium.},
Omer Nivron\footnote{University of Cambridge, UK.}, Enrique Portales Julia\footnote{University of Valencia, Spain.}, 
\\
Dietmar Backes\footnote{University of Luxembourg, Luxembourg.}, Piotr Bilinski\footnote{University of Oxford, UK.}, Guy Schumann\footnote{RSS-Hydro/University of Bristol, Luxembourg/UK.}.
\\
}

\begin{document}

\maketitle

\begin{abstract}
Reliable prediction of river floods in the first 72 hours can reduce harm because emergency agencies have sufficient time to prepare and deploy for help at the scene. Such river flood prediction models already exist and perform relatively well in most high-income countries. But, due to the limited availability of data, these models are lacking in low-income countries. Here, we offer the first global river flood prediction framework based on the newly published Caravan dataset. Our framework aims to serve as a benchmark for future global river flood prediction research. To support generlizability claims we include custom data evaluation splits. Further, we propose and evaluate a novel two-path LSTM architecture (2P-LSTM) against three baseline models. Finally,
we evaluate the generated models on different locations in Africa and Asia that were not part of the Caravan dataset.

\end{abstract}


\section{Introduction}

Floods can be devastating for humans, economies and nature. Recent catastrophic flooding events in Pakistan and Australia have highlighted the increasing severity of catastrophic flooding events. Between 1995 and 2015, over 2.6 billion people were affected by floods, accounting for 56\% of all people exposed to weather-related disasters \cite{HCWRD, vereinte_nationen_making_2015}. The impact of floods is worse in low and middle-income countries, which are home to 89\% of the world's flood-exposed population \cite{Rentschler2022FloodEA}. 

\par
Informed rescue operations, crisis management and planning can reduce the harm to the effected areas by floods. For example, people who receive flood warnings before the flood event are twice as likely to evacuate \cite{49993}. 
For such warning systems to have the desired impact, the UN highlights the importance of the first 72 hours after an event occurs \cite{unocha}.  
Flood forecasting systems with 72-hour lead-time currently only exist at a local level, mostly in high-income countries. 
This may be due to lack of globally available data and the complexity associated with hydrological models. 

\par
In this paper we offer the first global data-driven river flood prediction framework. This is made possible due to the newly published Caravan dataset, which is a global dataset that includes meteorological forcing data, catchment attributes and discharge data for basins around the world \cite{kratzert_frederik_2022_6578598}.


Our framework provides an end-to-end solution to train and test machine learning models for river flood predictions on a global scale. This is the first generalizable global flood prediction framework based on the Caravan dataset, to the best of our knowledge. The main contributions of this paper are:
(i) a framework to train and evaluate a global flood prediction model. The framework evaluates the generalizability of the models based on a test dataset defined using evaluation splits, i.e., if the predictions are generalizable to basins within the Caravans dataset, as well as to worldwide gauge locations that are not included in the Caravans dataset.
(ii) we propose and evaluate a novel model architecture (2P-LSTM) for river flood prediction, and,
(iii) Design experiments to evaluate the framework, train models using real-world Caravans data, and compare the 2P-LSTM model against three baselines which it outperforms for 72-hour ahead predictions. Furthermore, we extrapolate our predictions in basin areas located in low-income, not well-documented countries that were not part of the Caravans dataset.
Code for the framework and trained models are provided in {GitHub} (\url{https://github.com/mlahariya/floodcastai}). Python was used for the models developed in this article.
\par

\subsection{Application context}

\par
Why do humanitarian aid and disaster response agencies need a 72-hour notice of river-floods over the globe? The first 72 hours after a natural disaster occurs are crucial, according to the UN Office for the Coordination of Humanitarian Affairs (OCHA). 
Response must begin during that time to save lives. The first few hours are dedicated to assess the impact of the disaster. Then, OCHA deploys skilled staff to the scene, releases inter-agency appeal within 24-72 hours, and mobilizes initial funding within 72 hours \cite{unocha}. In short, if we are able to predict 72 hours before an event, disaster response agencies can initiate their 72-hour plans before the negative impact has already started.         

\par
Our research provides the first framework for global river flood predictions with up to 72-hours notice before the start of a flooding event.





\section{Data}
In this work we use the newly published Caravan dataset, which is a global 
dataset of 2532 gauge stations in river basins and their attributes (Figure \ref{fig:locations} shows the gauge locations). These attributes can be put in two categories: dynamic attributes and static attributes\footnote{For the full list of attributes refer to Kratzert et al.\cite{kratzert_frederik_2022_6578598}}. Dynamic attributes are measurements that change over time such as the daily temperatures, precipitation, etc. Static attributes are measurements that are considered to be constant such as the elevation of the river basin. 
\par
The framework we developed aims to identify when the water level tops the river bank given previous dynamic and static attributes. The proxy for water level in the Caravan dataset is the streamflow variable (mm/day) --- which describes the volume of water flowing through a river section divided by the area of the basin.   

\par
The Caravan dataset $\cal D$ is structured in the form of a daily time-series, one for each river basin. $t$ is on a daily granularity and $t \in [1980, 2022]$. For example, $t=1998.0026$ describes January 2nd 1998.
The target variable for daily mean streamflow (mm/day) at time $t$ will be denoted by $y_{t}$, where $y_{t} \in \mathbb{R}$. Furthermore, we denote the dynamic attributes (e.g. precipitation) at time $t$ as $x_t^{d}$, where $x_t^{d} \in \mathbb{R}^{40}$ and the static attributes (e.g. \% vegetation cover) of the specific basin which we denote by $x^s$, where $x^s \in \mathbb{R}^{209}$. 

\par
To ensure the model predictions can generalize to worldwide gauge locations that are not included in the Caravans dataset, we have curated a new independent dataset $\cal R$ of 195 river basins located in low and middle income countries are not represented that are not represented on Caravan. 
The data sources for our new dataset are: Global Runoff-Data Center \citep{GRDC} for our target streamflow measurements, the ERA5 project \citep{munoz2021era5} from the European Centre for Medium-Range Weather Forecasts (ECMWF) for our dynamic attributes, and HydroATLAS \citep{linke2019global} for the static attributes. These are the same sources that have been used to create the Caravan dataset. 


\section{Methodology}
\begin{figure}[!th]
\centering
\begin{minipage}[b]{.75\textwidth}
  \centering
  \includegraphics[width=1\textwidth]{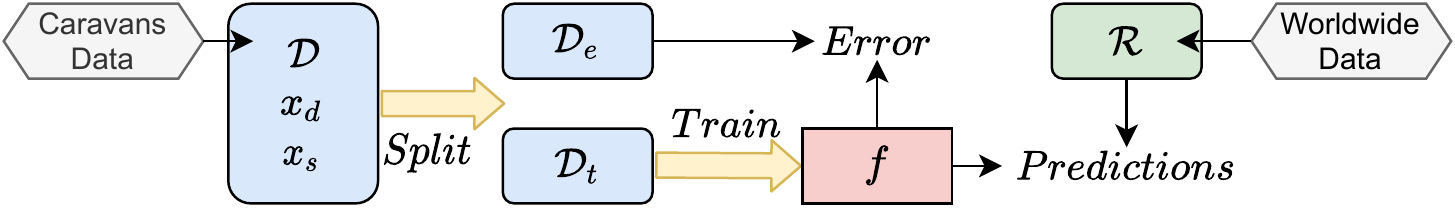}
  \caption{Overview of the framework to train model \emph{f}.}
  \label{fig:ml-diagram}
\end{minipage}
\end{figure}

In this section, we describe the framework used to train and evaluate a global flood prediction model. The machine learning pipeline is shown in Figure~\ref{fig:ml-diagram}.
\subsection{Models}
Our machine learning pipeline focuses on the daily mean streamflow prediction at each gauge station. We develop a framework that predicts the streamflow at each time step $t, t+1, t+2, t+3$, as to achieve 72-hour lead-time predictions. Following the mathematical notation introduced in section 2, the learning setting is:

\begin{equation}
\hat{y}_{t:t+3} = f(x_t^{d}, \dots, x_{t-k}^{d}, x^{s}, y_{t-1}),
\end{equation}

where $f$ denotes each model written as a function, mapping the input variables to the predicted streamflow sequence $\hat{y}_{t:t+3}$ and $k$ is the length of the look-back sequence. Since the temporal resolution of the data is daily, each time step $t+m$ represents $m$ days-ahead quantities. The estimated sequence $\hat{y}_{t:t+3}$ depends on the last $k$ dynamic variables, meaning $\{x^d_{t}, ... x^d_{t-k} \}$, the static variables $x^s$ and streamflow value at the previous time step $y_{t-1}$. To extend the prediction task to a 72-hour lead-time, we estimate $\hat{y}_{t}, \hat{y}_{t+1}, \hat{y}_{t+2}, \hat{y}_{t+3}$.

\paragraph{Two-path architecture (2P-LSTM)} The 2P-LSTM architecture focuses on modelling separately the dynamic and static attributes. Such an approach can be effective when modelling hydrological data that often consists of dynamic variables (e.g. hydrological signatures) and static attributes (e.g. catchment properties).
Separate modelling of static and dynamic attributes has been previously studied in recent literature and has provided promising results in the hydrological domain \cite{leontjeva2016, anshuman2022,kratzert2019}.
 
\begin{figure}[!th]
\centering
\begin{minipage}[b]{\textwidth}
  \centering
  \includegraphics[width=1\textwidth]{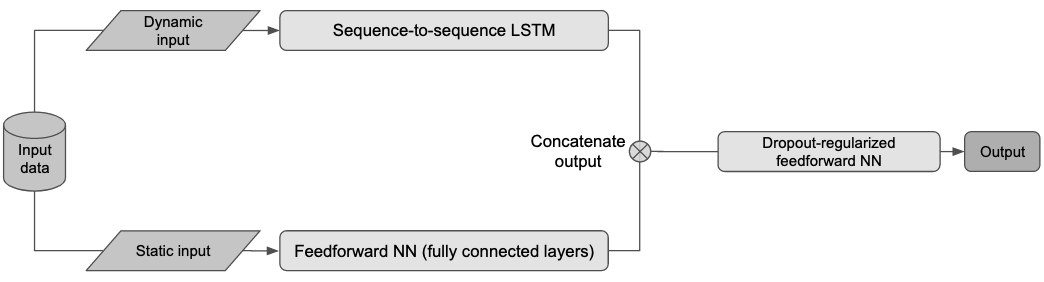}
  \caption{An overview of the 2P-LSTM architecture.}
  \label{fig:2PLSTM_arch}
\end{minipage}%
\end{figure}

As it is illustrated in figure \ref{fig:2PLSTM_arch}, the 2P-LSTM architecture separates the input data into two modelling paths. The dynamic variables follow a path through a sequence-to-sequence LSTM network (Seq2Seq LSTM), and the static attributes follow a path through a feedforward network \cite{sutskever2014}. The outputs of the two paths are then concatenated and serve as input to a feedforward network. Finally, the output of the network consists of a sequence of predicted streamflow values over the time steps $t, t+1, t+2, t+3$.

The Seq2Seq LSTM that models the dynamic attributes consists of two networks: an encoder and a decoder. The encoder learns to predict a streamflow value using the time-dependent features in the look-back sequence $k$. The outputs of the encoder are discarded and its cell state and hidden state information are used to initialize the decoder network. The decoder then generates the output sequence predicting the streamflow values in the next time step. In training mode, the decoder takes as input a streamflow sequence of $y_{t-1}, y_t, y_{t+1}, y_{t+2}$, and predicts an estimated streamflow sequence $\hat{y}_t, \hat{y}_{t+1}, \hat{y}_{t+2}, \hat{y}_{t+3}$. However, in inference mode, the input streamflow sequences are not available. In this case, only the streamflow value $y_{t-1}$ is used as input, and, in a recursive manner, the resulting estimate for $\hat{y}_t$ is used as input to predict the streamflow at the next time step.

We compare the proposed 2P-LSTM with three baseline models: linear regression (LR), random forest (RF) ensembles, and a feedforward neural network (NN).

\subsection{Evaluation}

A generalizable model should be able to make predictions in ungauged basins and in locations where there is no training data. Additionally, the model should generalize to unrepresented climates zones.
 Traditional model validation methods (such as random k-fold validation) are wanting in appropriate estimate of models performance on ungauged basins and unrepresented climates -- to ensure its generalizability globally.
To ensure models global generalizability, we define custom \emph{Evaluation splits} on  split Caravan dataset into train, validation and test datasets. These test datasets are prepared based on two types of splits, namely, (i) holdout split, and (ii) stratified split.

In \emph{Holdout split}, the test set contains the data from ungauged basin in unrepresented climates. Thus, historical information of the gauges and climate zones in the test set is not used to train the model.
In \emph{Stratified split}, the test set contains the data from ungauged basins in seen climates. For each climate zone, we keep 10\% of the gauges in the test set, 10\% in the validation set, and 80\% in the training set. Thus, we do stratified sampling of gauges based on climate zones based on the climate zone in the basin area. 



\section{Experimental Results}


Tables \ref{rmsestrat} and \ref{rmseholdout} report the RMSE, obtained from the stratified and holdout test sets respectively. First, notice that the errors for the stratified split are $\sim$ 1 mm/day for $y_t$. The predictions for $y_{t+1}, y_{t+2}, y_{t+3}$ show bigger errors, as expected due to cumulative time-series error. Second, the errors for the holdout split are slightly higher. The prediction task for the holdout split is harder because the test set contains river basins from unrepresented climate zones.


\begin{table}[h!]
    \begin{minipage}{.47\linewidth}
        \setlength{\tabcolsep}{2mm}
        \caption{RMSE for models on \emph{stratified split} test dataset}
        \label{rmsestrat}
        \centering
        \begin{tabular}{@{}llllll@{}}
        \toprule[2pt]
        
        & \multicolumn{4}{c}{Time-step} \\
        \cmidrule(lr){2-5} \emph{(mm/day)}
           & y$_{t}$                        & y$_{t+1}$                          & y$_{t+2}$                           & y$_{t+3}$                           &  \\ \midrule
        LR      & 1.81                        & 2.10                        & 2.45                        & 2.55                        &  \\
        RF      & \textbf{ 1.42 }& \textbf{1.76} & 2.07 &  2.23 &  \\
        NN    & 1.86                        & 2.46                        & 2.70                        & 2.84                        &  \\
        2P-LSTM & 1.43                        & 1.86                        & \textbf{1.97}                        & \textbf{2.13}                        &  \\ 
        \bottomrule[2pt]
        \end{tabular}
    \end{minipage}
    \hspace{10pt}
    \begin{minipage}{.47\linewidth}
        \setlength{\tabcolsep}{2mm}
        \caption{RMSE for models on \emph{holdout split} test dataset   }
        \label{rmseholdout}
        \centering
        \begin{tabular}{@{}llllll@{}}
        \toprule[2pt]
        
        & \multicolumn{4}{c}{Time-step} \\
        \cmidrule(lr){2-5} \emph{(mm/day)}
           & y$_{t}$                        & y$_{t+1}$                          & y$_{t+2}$                           & y$_{t+3}$                           &  \\ \midrule
        LR      & 2.44                        & 3.18                        & 3.55                        & 3.55                        &  \\
        RF      &  1.64 &  \textbf{1.92 }& \textbf{2.18} &  \textbf{2.31} &  \\
        NN    &  2.10 &  2.69 &  2.89 & 3.00 &  \\
        2P-LSTM & \textbf{1.58 }                       & 1.94                        & 2.25                        & 2.49                        &  \\
        \bottomrule[2pt]
        \end{tabular}
    \end{minipage}
\end{table}

\begin{figure}[!th]
\centering
\begin{minipage}[b]{0.8\textwidth}
  \centering
  \includegraphics[width=1\textwidth]{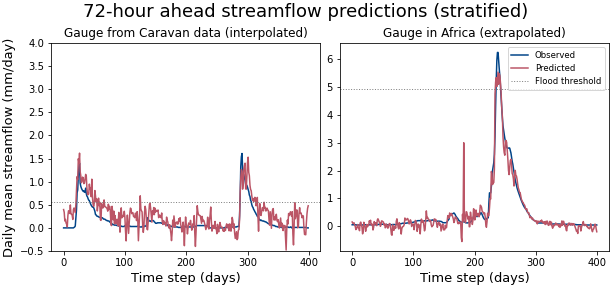}
  \caption{72-hour ahead streamflow predictions produced by the 2P-LSTM model trained using the stratified split. On the right, predictions are extrapolated to a gauge in Africa, outside the Caravan data.}
  \label{fig:2PLSTM_strat}
\end{minipage}%
\end{figure}

With respect to the models, RF and 2P-LSTM are the best performing ones, with similar results for both stratified and holdout test splits. However, 2P-LSTM yields the lowest error in predicting $y_{t+3}$. Since our aim is to have the best model for 72-hours ahead predictions, we chose 2P-LSTM to illustrate some predictions. Figure \ref{fig:2PLSTM_strat} shows a comparison between observed and predicted streamflow for two cases: one interpolated case where the test gauge belongs to Caravan data (left) and an extrapolated case, where the test gauge belongs to the non-Caravan dataset from Africa (right). 
In most cases, the model accurately captures the location and magnitude of the high flow peaks, which are possible cases of  river overflow. In this non-Caravan test set, the 2P-LSTM evaluation error was 2.59 mm/day, which is an example of accurate predictions and generalization capabilities considering that these basins are unrepresented in training. 


\section{Conclusions}

In this paper, we propose the first global river flood prediction framework based on the newly published Caravan dataset. Our framework aims to serve as a benchmark for future global river flood prediction research. We show the high potential of our framework by learning and evaluating three baseline models and a novel 2P-LSTM architecture. Based on the evaluation, the 2P-LSTM model outperforms the baseline models, and provides 2.13 mm/day RMSE in the gauges in Caravan dataset and 2.59 mm/day RMSE when extrapolating to worldwide gauges outside of the Caravans dataset. This is an important step towards river flood predictions in low-income, poorly-documented areas around the world. Future work includes, (i) flood extent derivation using the predicted streamflow in combination with Digital Elevation Models (DEMs), and,
(ii) evaluating the potential of other sources of data in model inference, such as forecasted data and satellite data.




\small
\bibliography{references.bib}
\newpage
\appendix

\section{Appendix}

\begin{figure}[!th]
\centering
\begin{minipage}[b]{.7\textwidth}
  \centering
  \includegraphics[width=1\textwidth]{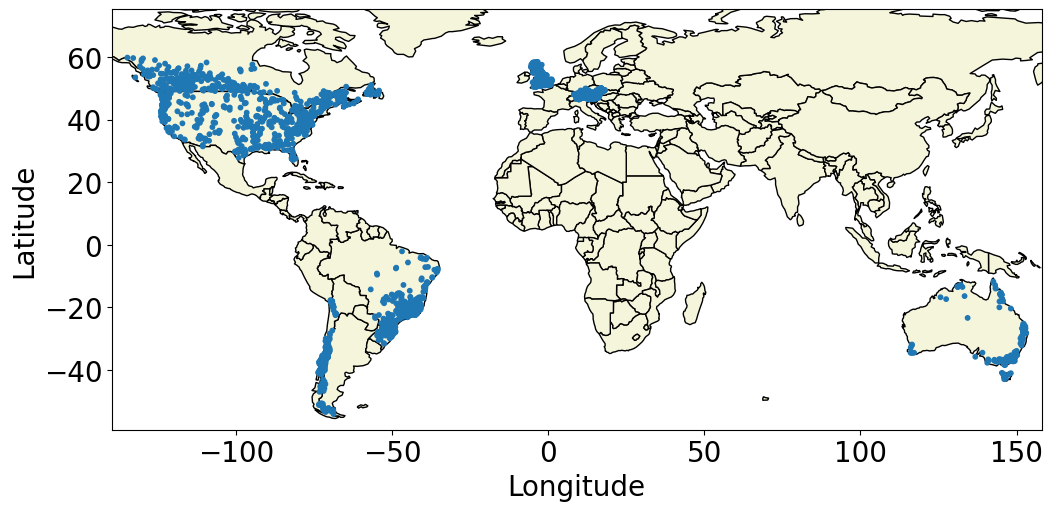}
  \caption{Locations of the basins contained in Caravans dataset.}
  \label{fig:locations}
\end{minipage}%
\end{figure}

\begin{figure}[!th]
\centering
\begin{minipage}[b]{0.8\textwidth}
  \centering
  \includegraphics[width=1\textwidth]{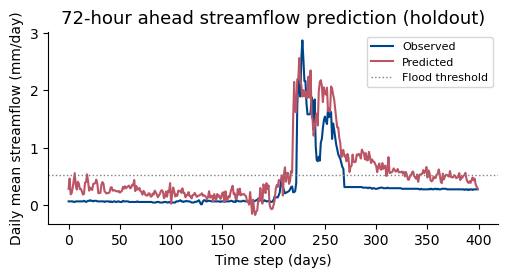}
  \caption{72-hour ahead streamflow predictions produced by the 2P-LSTM model trained using the holdout split.}
  \label{fig:2PLSTM_ho}
\end{minipage}%
\end{figure}

\end{document}